\documentclass{PoS}
\usepackage{amsmath}

\title{Minimal Model for an Unbalanced Holographic Superconductor}

\ShortTitle{Minimal Model for an Unbalanced Holographic Superconductor}

\author{\speaker{Daniele Musso}%
         \\
        $\cdot$ Physique Th\'eorique et Math\'ematique\\
        Universit\'e Libre de Bruxelles, C.P. 231, 1050, Bruxelles, Belgium\\
        $\cdot$ International Solvay Institutes, Brussels, Belgium  \\      
        E-mail: \email{dmusso@ulb.ac.be}}

\abstract{We describe the simplest holographic model
for an s-wave unbalanced superconductor in $2+1$ dimensions. 
We study its phase diagram and linear response features with particular
attention to the possibility of spatially modulated phases (LOFF) 
and mixed spin-electric properties.
The normal phase of the model at hand allows us to analyze a 
strong-coupling generalization of Mott two-current model for 
spintronic systems; the superconducting phase features an interesting
DC spin-superconductivity without spin-symmetry breaking .}

\FullConference{ Proceedings of the Corfu Summer Institute 2012 \\
                 September 8-27, 2012\\
                 Corfu, Greece}

\begin{document}

\section{Introduction}

\subsection{Theoretical framework}

The dynamics of quantum field theory at strong coupling has always
represented a tough subject to be studied. Especially because it is troublesome
to access the strongly coupled regime of QFT by means of standard methods,
generally based on perturbation theory. 
Even adopting numerical lattice methods, the analysis of
finite density systems and transport properties is in general particularly
delicate and difficult.
However, in the last fifteen years, new theoretical insight inspired by 
string theory and brane dynamics has opened a new theoretical path to the 
quantitative study of QFT at strong coupling.
Such an alternative approach is referred to with the term \emph{holography} as
it is based on a conjectured duality between specific examples of QFT and 
suitable gravitational models living in a spacetime with higher dimensionality.
The paradigmatic example of holographic correspondence is the $AdS$/CFT 
\cite{Aharony:1999ti,Zaffaroni:2005ty}.

Holographic dualities conjecture a correspondence between
a QFT (without gravity) and a gravitational model; 
the former is generally thought of as living on the ``boundary''
of a bulk manifold where instead the dual gravitational model is defined.
The practical power of holography descends from the fact that
the strongly coupled regime of the QFT is mapped to the semiclassical
regime of the dual gravity model. In other terms, a semiclassical approach 
based on the direct solution of the equations of motion on the gravity side
allows one to obtain quantitative information about the correlation functions
of the boundary field theory.

The holographic framework, or gauge/gravity correspondence, has been widely 
and deeply considered for a plethora of applications like the physics of 
strongly coupled plasmas and the quantum phase transitions. 
A more specific but nevertheless very important field of application 
is the one concerning holographic superconductors \cite{Gubser:2008px,Hartnoll:2008vx,Hartnoll:2008kx}%
\footnote{For an ampler perspective consult the review article \cite{Horowitz:2010gk,Herzog:2009xv,Hartnoll:2009sz}.}.
The holographic superconductors are strongly coupled systems manifesting symmetry 
breaking phenomena which lead to superconductivity; they are useful
toy models to study unconventional (non-BCS) superconductivity and possibly
shed light on some features of the high-$T_c$ superconductivity mechanisms.
They implement in a holographic context the essential ingredients to describe 
the breaking of an Abelian symmetry and the consequent superconducting phenomenon;
the minimal model contains a bulk vector gauge field and a charged scalar living
in a black hole (i.e. finite temperature) background. The condensation of the 
charged scalar, namely the black hole hair formation, gives rise to a bulk
Higgs mechanism. From the dual standpoint, this condensation is interpreted as the phase transition
leading to a strongly coupled superfluid whose one and two-point correlation functions
coincide with those of a strongly coupled superconductor.

\subsection{Motivations}

In \cite{Bigazzi:2011ak} we extended the minimal model 
for a holographic superconductor \cite{Hartnoll:2008kx} by introducing a second
$U(1)$ gauge field in the bulk theory.
This generalization aims at describing a boundary system with
two chemical species associated with two chemical potentials.
Indeed, following the so-called holographic dictionary, a gauge symmetry in the bulk
is related to a conserved current in the boundary and, more specifically,
the boundary condition for the time component of the bulk gauge field is 
interpreted as a chemical potential in the boundary theory.

Having a system with two chemical potentials allows us to study 
the so-called unbalanced systems characterized by the presence of a chemical
imbalance between different chemical species%
\footnote{An earlier study of unbalanced, holographic systems has been proposed
in \cite{Erdmenger:2011hp}.}. This class of systems 
is extremely important in a wide variety of 
physical situations ranging from the QCD to the condensed matter panorama.
Listing just a few examples, we have: cold atoms, 
neutron stars and unconventional unbalanced superconductors.
The holographic approach aims to study unbalanced mixtures at strong coupling
and allows us to handle both equilibrium or slightly out of equilibrium features,
that is to say, both the thermodynamics and the transport properties.

Even though many crucial theoretical steps in the realm of weakly coupled unbalanced systems
have been attained in the early days of the BCS theory%
\footnote{For a review of the standard BCS, weak-coupling treatment of unbalanced superconductors consult \cite{casnar}.}, the experimental
technology has started to allow us to investigate directly many of their properties only in recent times.
In particular, to investigate phenomena like LOFF phases (see next Subsection),
very stringent conditions and especially low spin relaxation rates are required.
Furthermore, both theoretically and experimentally, it is crucial to understand whether and
how the weak coupling picture changes at strong coupling. 
This is one of the main purposes of the present holographic analysis.

\subsection{Weak coupling picture}

The weak coupling description of Fermi systems is captured by the Fermi surface physics
and the corresponding quasi-particle excitations.
At low temperature, due to the Pauli exclusion principle, Fermions ``pile up'' progressively occupying
higher energy levels up to the Fermi surface.
In the presence of more than one Fermionic chemical species,
each species gives rise to a Fermi surface. When the Fermi levels
for distinct chemical species have different values, the system is said to be
\emph{unbalanced}. 

Among the unbalanced systems, the unbalanced superconductors are particularly 
interesting. Here the chemical species are two, namely ``spin-up'' and
``spin-down'' electrons. The chemical imbalance in such a situation can 
be produced by the presence of magnetic impurities in the system or, 
for instance, an external magnetic field inducing Zeeman splitting of single-electron
energy levels%
\footnote{We remind the reader that even though we mainly stick to the 
condensed matter applications and language the holographic approach furnishes
interesting simple models for QCD applications as well. Hence, even when 
the treatment specializes, one should not forget the flexibility of the original
toy model.}.
Already at weak coupling, through a BCS analysis, it is possible
to uncover interesting phenomena like the occurrence of inhomogeneous phases
where the superconducting condensate acquires spontaneously non-trivial
spatial modulations \cite{loff}. Such exotic phases with spatially modulated
condensates are called LOFF; even though predicted theoretically,
these LOFF phases have not yet been definitively confirmed by experiments.
%\footnote{A promising field for LOFF phases experimental investigations 
%is furnished by cold atoms (WARNING).}.

\begin{figure}[t]
\centering
\includegraphics[width=100mm]{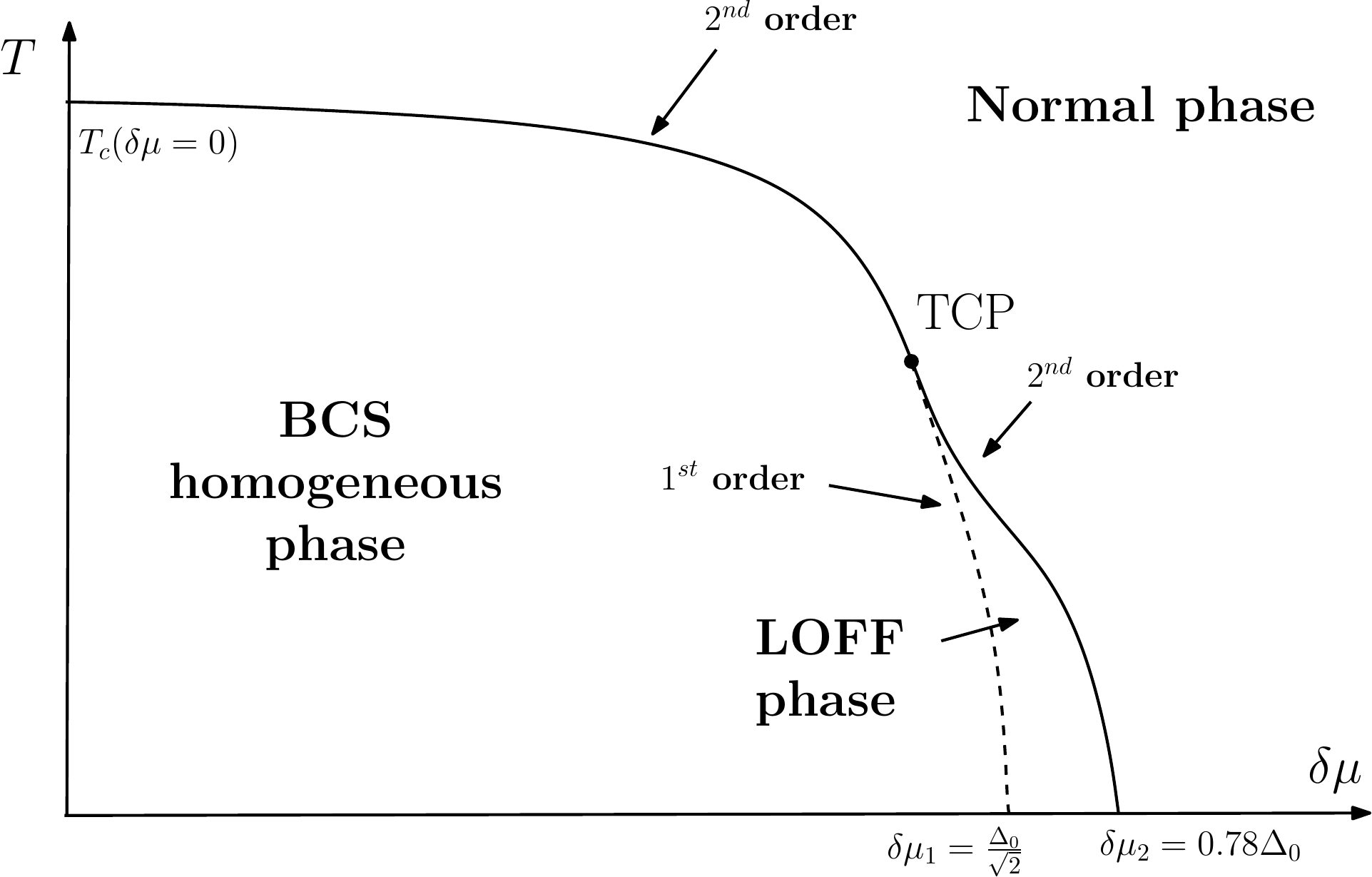}
\caption{Phase diagram for a weakly coupled, unbalanced superconductor.
The Chandrasekhar-Clogston bound is $\delta\mu_1$; above such bound, 
inhomogeneous superconductivity is thermodynamically favored. 
Eventually, for $\mu > \delta\mu_2$, superconductivity is lost.
The solid lines represent second order phase transitions while the dashed 
line correspond to first order phase transitions; the lines meet at a tricritical point.
$\Delta_0$ is the gap parameter of the balanced superconductor (i.e. $\delta\mu=0$) at zero temperature
(for further details consult \cite{casnar}).}
\label{weak_phase}
\end{figure}

Within a BCS treatment, superconductivity is associated with the so-called
Cooper pairing mechanism due to a phonon mediated attractive interaction
between electrons. More specifically, in an s-wave superconductor, 
the two electrons forming a Cooper pair have oppositely oriented spin,
the pair as a whole is then in an s-wave state.
It is therefore intuitive that a chemical imbalance between ``spin-up'' and
``spin-down'' electrons is likely to hinder the s-wave Cooper pair mechanism.
Indeed, at weak coupling, it is possible to predict the existence of a maximal 
value for the imbalance above which homogeneous superconductivity is lost;
such limiting value for the chemical imbalance is called Chandrasekhar-Clogston
bound \cite{cc}, see Figure \ref{weak_phase}.
Above the Chandrasekhar-Clogston bound, the homogeneous superconducting phase
becomes thermodynamically disfavored with respect to spatially modulated phases. 
Microscopically, a spatial modulation for the Cooper condensate corresponds 
to having Cooper pairs with a finite center of mass momentum.

Another very interesting aspect of unbalanced systems is the occurrence of
mixed transport phenomena. Sticking to the superconductor example,
we have mixed spin-electric transport features, in one word, ``spintronics''%
\footnote{For an introductory and general treatment of spintronics see \cite{spintronics}.}. 
It is not difficult to observe that a chemical imbalance leads to mixed spin-electric behavior:
consider a material having an itinerant cloud of electrons whose spins are prevalently oriented
along the ``up'' direction; an external electrical perturbation will of course induce an electric 
response (i.e. electric transport) and, at the same time, also spin transport. In other words,
there is a net magnetic transport in response to an electrical perturbation, the converse being true as well.
Apart from their very important technological applications, it is interesting
to study spintronic behaviors in holographic systems where a possible strong coupling realization
is accessed.

\section{A holographic model}

\subsection{Comment on the holographic ``effective'' approach to study superconductivity}

Many central features which represent the hallmark of superconductivity,
such as a diverging DC conductivity and the Meissner-Ochsenfeld effect,
are a direct consequence of the spontaneous symmetry breaking \cite{Weinberg:1986cq}.
The $U(1)$ symmetry breaking is therefore the crucial ingredient of any 
theoretical model aiming to reproduce the superconductor phenomenology, both at weak
and strong coupling. Indeed, both standard Ginzburg-Landau approaches and the
holographic approaches comply to this paradigm.

It is nevertheless useful to pinpoint some differences between the standard effective 
theories and the $AdS$/CFT inspired ones. The crucial distinction relies of course
on the fact that in holography we adopt a dual perspective. The gravitational side
of the duality is, or at least is assumed to be, the effective low-energy theory of
a UV complete theory in a standard sense; nevertheless, the gravitational effective theory
represents the strongly coupled system from a dual standpoint. 
Indeed, the gravitational low-energy fields correspond to dual gauge invariant
operators in the boundary model which, in principle, can receive contributions from 
all the modes of the boundary theory. In other words, no UV cutoff is in general considered 
in the boundary theory, which, in terms of the gravity model, corresponds to the fact that the radial 
bulk coordinate is considered up to infinity.

Another crucial point at the foundation of the holographic effective approach is the
large $N$ limit; it is only in this limit of a large number of degrees of freedom
that the strongly coupled boundary theory admits a dual, semiclassical description. 
The coarse graining procedure of standard effective field theory could remind
us about taking into account many degrees of freedom collectively by using
only a small number of effective fields. However, the coarse graining idea is again 
related to integrating out high energy modes while in holography the large $N$ hypothesis
is related to the possibility of accounting for strongly coupled quantum dynamics in a dual
classical perspective without implying any energy cutoff.

As an example of how the intrinsic large $N$ character of holography manifests itself,
let us consider the scalar condensation leading to the holographic superconducting phase.
From the bulk perspective, the scalar potential has only a (negative) quadratic mass term 
(as opposed to the standard Ginzburg-Landau quartic potential); we do not need
to cure the unboundedness from below of the scalar potential as the gravitational
interactions do the job for us. We have to remind ourselves that the gravitational 
interactions, or equivalently the geometry, are the semiclassical dual features accounting 
for a large number of degrees of freedom. Indeed, the 
scalar hair condensation occurs as the bulk scalar field crosses the infrared BF bound
and this is dual to the contemporary condensation of a large $N$ number of degrees 
of freedom in the boundary theory.

\subsection{Action and equations of motion}

We generalize the standard model for a holographic superconductor 
described in \cite{Hartnoll:2008kx} and introduce a second gauge field. 
The bulk action for such a generalized model is
\begin{equation}\label{zione}
 S = \frac{1}{2\kappa_4^2} \int dx^4 \sqrt{-g}
 \left[{\cal R} + \frac{6}{L^2} 
 - \frac{1}{4} F_{ab}F^{ab}
 - \frac{1}{4} Y_{ab}Y^{ab}
 - V(|\psi|)
 - |\partial \psi - i q A \psi |^2 \right] \ ,
\end{equation}
where $F=dA$ and $Y=dB$ are the two field strengths associated to the two
gauge fields; note that the scalar $\psi$ is charged
only under the ``electric'' gauge field $A$. 
With the aim of studying some particularly simple and useful solutions
of the equations of motion descending from \eqref{zione},
we consider the following standard ansatz
\begin{equation}\label{an}
 ds^2 = -g(r) e^{-\chi(r)} dt^2
 + \frac{r^2}{L^2} (dx^2 + dy^2)
 + \frac{dr^2}{g(r)}\ ,
\end{equation}
\begin{equation}\label{satz}
 \psi = \psi(r)\ , \ \ \ \
 A_a\, dx^a = \phi(r)\, dt\ , \ \ \ \
 B_a\, dx^a = v(r)\, dt\ .
\end{equation}
All the fields have only radial dependence and are constant with respect
to the remaining coordinates parameterizing the boundary manifold.
Furthermore, since one of the Maxwell equations implies that the phase of $\psi$ is constant,
we take it to be null; this corresponds to consider $\psi$ to be real.
We henceforth choose units in which $L=1$ and $2\kappa_4^2=1$.
Employing the ansatz \eqref{an} and \eqref{satz}, the equations of motion
get the following explicit form
\begin{equation}
 \psi'' + \psi' \left(\frac{g'}{g} + \frac{2}{r} - \frac{\chi'}{2}\right)
 - \frac{V'(\psi)}{2 g}
 + \frac{e^\chi q^2 \phi^2 \psi}{g^2} = 0
\end{equation}
\begin{equation}
 \phi'' + \phi'\left( \frac{2}{r} + \frac{\chi'}{2} \right)
 - \frac{2 q^2 \psi^2}{g} \phi = 0
\end{equation}
\begin{equation}
 \frac{1}{2} \psi'^2 
 + \frac{e^\chi (\phi'^2 + v'^2)}{4g}
 + \frac{g'}{gr}
 +\frac{1}{r^2}
 -\frac{3}{g}
 + \frac{V(\psi)}{2g}
 + \frac{e^\chi q^2 \psi^2 \phi^2}{2g^2} = 0
\end{equation}
\begin{equation}
 \chi' + r\psi'^2
 + r \frac{e^\chi q^2 \phi^2 \psi^2}{g^2} = 0
\end{equation}
\begin{equation}
 v'' + v'\left( \frac{2}{r} + \frac{\chi'}{2} \right) = 0
\end{equation}

We actually specialize the treatment and consider $V(\psi)= m^2 \psi^2$
with $m^2=-2$. This mass choice is standard as it arises from many 
consistent truncations of string theory and supergravity \cite{Gauntlett:2009bh,Bobev:2011rv}.
The explicit mass choice affects the large $r$ asymptotic
behavior of the scalar field, in the case at hand we have
\begin{equation}
 \psi(r) = \frac{C_1}{r} + \frac{C_2}{r^2} + ...
\end{equation}
We interpret the coefficient of the near-boundary leading term $C_1$ as
corresponding to the dual source of the operator whose 1-point expectation
value is accounted for by $C_2$%
\footnote{The value of mass considered explicitly, $m^2=-2$, falls within
the interval where two quantizations for the scalar field on $AdS_4$ are possible; 
roughly speaking, this means that we could have interpreted $C_2$ as the source and $C_1$ 
as the expectation of the corresponding operator. The two quantizations differ 
by boundary terms that determine which boundary field theory we are
studying holographically \cite{Klebanov:1999tb}.}.
Without entering into further technical detail, we will consider $C_1=0$ according to the fact
that we want to study unsourced, i.e. spontaneous, condensation of the corresponding operator $\cal O$,
\begin{equation}
 C_1 = 0\ , \ \ \ \
 \langle {\cal O} \rangle = \sqrt{2}\ C_2\ .
\end{equation}
The conventional factor $\sqrt{2}$ is introduced to agree with the existing literature.

The asymptotic near-boundary behavior of the gauge fields is
\begin{equation}
 \phi(r) = \mu - \frac{\rho}{r} + ... \ , \ \ \ \ 
 v(r) = \delta\mu - \frac{\delta\rho}{r} + ... 
\end{equation}
where, from the boundary theory standpoint,
the leading terms are interpreted respectively as chemical potentials
and charge densities. These are standard entries of the holographic
dictionary. Eventually, requiring regularity of the Euclidean time at the horizon, 
we obtain the following expression for the black hole temperature, 
\begin{equation}
 T = \frac{r_H}{16 \pi} \left[
 \left(12 - 2 m^2 \psi^2_{H0}\right) e^{-\frac{\chi_{H0}}{2}}
 - \frac{1}{r_H^2} e^{\frac{\chi_{H0}}{2}} \left(\phi^2_{H1} + v^2_{H1}\right)\right]\ ,
\end{equation}
where $r_H$ is the horizon radius; the subindex $H0$ refers to constant terms at the
horizon while $H1$ refers to first order terms in $r-r_H$.
Both the bulk and the boundary theories have the same time coordinate and, consequently,
also the same complex time continuation and temperature.

\section{Brief account of the results}

\subsection{Equilibrium}

\begin{figure}[t]
\centering
\includegraphics[width=70mm]{phastrong.pdf} \hspace{0.5cm}
\includegraphics[width=70mm]{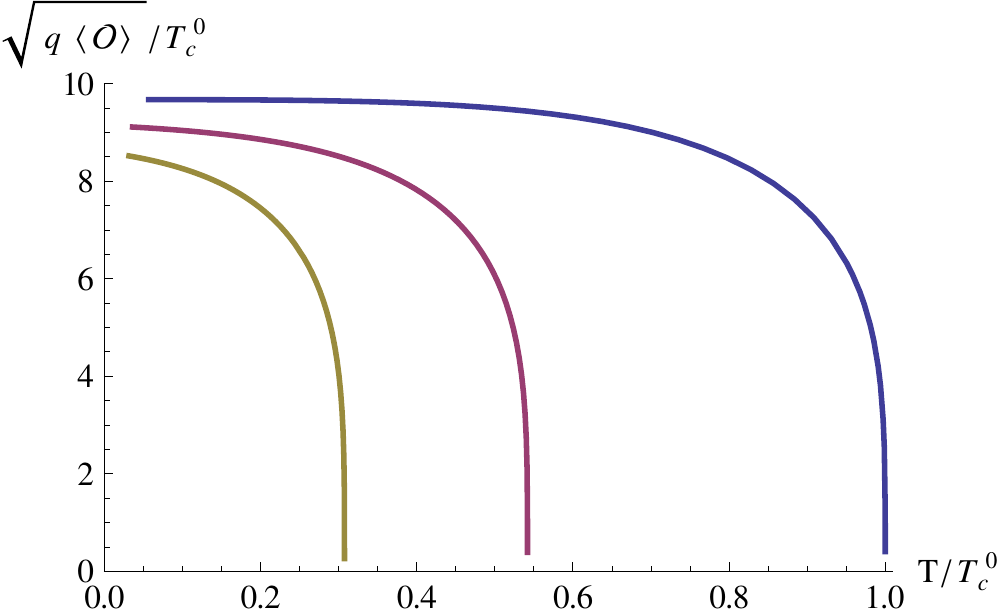}
\caption{On the left we have the phase diagram of unbalanced holographic
superconductor described in the main text.
Notice the absence of the Chandrasekhar-Clogston bound (see Figure %\ref{weak_phase}). 
On the right we have various scalar condensates plotted against temperature. 
The condensation occurs at lower temperature for increasing values of the
chemical imbalance $\delta\mu$.}
\label{conde}
\end{figure}

In Figure \ref{conde} we plot the $\psi$ condensates for $\delta\mu = 0, 1, 1.5$ respectively.
We observe that increasing the chemical imbalance, the superconducting condensation
occurs at a lower critical temperature.
This corresponds to the fact that the imbalance hinders the condensation;
such result obtained at strong coupling is in line with the weak coupling expectation.

On the other hand, the weakly coupled unbalanced superconductor presents a Chandrasekhar-Clogston
bound $\delta\mu_1$ beyond which homogeneous conductivity is lost (see Figure \ref{weak_phase}).
In our specific holographic model, we do not find any such bound; said the other way round,
for any value of $\delta\mu$ we have condensation at sufficiently low temperature.
Note however that this result could be sensitive to the details of the model and, in particular, to 
the values of the parameters like the scalar field mass.

\subsection{Linear response}
\label{LR}

One of the most interesting features of the holographic model at hand
relies on the possibility of studying its mixed spin-electric
linear response properties. In this sense, it is tempting to regard the model as a generalization
at strong coupling of the simplest spintronic models, namely the Mott two-current model and its
generalizations \cite{mott}.

The linear response of the system is described with the conductivity matrix
\begin{eqnarray}\label{matrix}
\begin{pmatrix}  J^A \\ Q \\ J^B \end{pmatrix} = \begin{pmatrix}  \sigma_A & \alpha T & \gamma \\ \alpha T &
\kappa T & \beta T \\ \gamma & \beta T & \sigma_B \end{pmatrix} \cdot \begin{pmatrix}  E^A \\ -\frac{\nabla T}{T} \\ E^B \end{pmatrix} \, .
\end{eqnarray}
which encodes ``electric'', ``spin'' and thermal response. The off-diagonal components are
obviously associated to mixed effects; for instance, $\alpha$ accounts for the ``thermo-electric'' 
response.

To study the transport behavior of our thermodynamical system we have to consider
small variations of the sources and the consequent current flows. Holographically,
this translates in studying the equations of motion for vector fluctuations on 
the fixed gravitational background, the latter corresponding to the thermodynamical 
equilibrium state of the boundary theory. To have a detailed account on how the 
physical quantities appearing in \eqref{matrix} are related to the dual gravitational 
fields (i.e. the holographic dictionary applied to fluctuation fields)
we refer to \cite{Bigazzi:2011ak,Musso:2012sn}.

Without any loss of generality, we choose the vector fluctuations
to be along the $x$ direction. In our model, the vector fluctuations 
involve the gauge fields and the vector mode of the metric, i.e. $g_{tx}$.
The system of equations of motion for the fluctuations is
\begin{equation}\label{maxa}
 A_x'' + \left(\frac{g'}{g}-\frac{\chi'}{2}\right) A'_x + \left(\frac{\omega^2}{g^2}e^\chi - \frac{2q^2\psi^2}{g}\right) A_x =
 \frac{\phi'}{g} e^\chi \left(-g'_{tx} + \frac{2}{r} g_{tx}\right)\, ,
\end{equation}
\begin{equation}\label{maxb}
 B_x'' + \left(\frac{g'}{g}-\frac{\chi'}{2}\right) B'_x + \frac{\omega^2}{g^2}e^\chi B_x =
 \frac{v'}{g} e^\chi \left(-g'_{tx} + \frac{2}{r} g_{tx}\right)\, ,
\end{equation}
\begin{equation}\label{grav}
 g'_{tx} - \frac{2}{r}g_{tx} + \phi' A_x + v' B_x= 0\, .
\end{equation}
Upon substituting \eqref{grav} into the other equations we obtain
\begin{equation}\label{eqA}
 A_x'' + \left(\frac{g'}{g} - \frac{\chi'}{2}\right) A_x'
 + \left( \frac{\omega^2}{g^2}e^{\chi} - \frac{2 q^2 \psi^2}{g} \right) A_x
 - \frac{\phi'}{g} e^{\chi} \left(B_x v' + A_x \phi' \right) = 0\, ,
\end{equation}
\begin{equation}\label{eqB}
 B_x'' + \left(\frac{g'}{g} - \frac{\chi'}{2}\right) B_x'
 + \frac{\omega^2}{g^2}e^{\chi} B_x
 - \frac{v'}{g} e^{\chi} \left(B_x v' + A_x \phi' \right) = 0\, .
\end{equation}
The step just performed is crucial: the metric fluctuations couple
the two equations for the gauge fields $A$ and $B$. This coupling gives rise to the 
mixed spin-electric transport properties of the system; since the metric is directly 
involved in coupling the equations for the fluctuations of $A$ and $B$, 
the mixed $A-B$ character of the system relies on considering the backreaction 
of the gauge fields on the geometry.
In addition, note that the metric vector fluctuations disappeared
from the equations of motion after having substituted \eqref{grav}.

The symmetry of the conductivity matrix \eqref{matrix} and the interpretation
of the second $U(1)$ as describing effectively magnetic degrees of freedom could be accommodated
considering appropriate time-reversal assignments for the fields of the system.
Indeed we could assume that the gauge field 
(or ``vector potential'') $B$ behaves oppositely with respect to $A$ under time reversal,
\begin{eqnarray}
 (A_t,A_i) &\longrightarrow& (A_t,-A_i)\\
 (B_t,B_i) &\longrightarrow& (-B_t,B_i)
\end{eqnarray}
so 
\begin{equation}\label{TR}
 \phi \rightarrow \phi\ , \ \ \ \
 A_x \rightarrow -A_x\ , \ \ \ \
 v \rightarrow -v\ , \ \ \ \
 B_x \rightarrow B_x\ , \ \ \ \
 g_{tx} \rightarrow -g_{tx}\ ,
\end{equation}
\begin{equation}
 \mu \rightarrow \mu \ , \ \ \ \ 
 \delta \mu \rightarrow -\delta \mu\ .
\end{equation}
Notice that the equations of motion for both the background and the fluctuations are
invariant under the transformation \eqref{TR}.

One of the most striking results which emerged is that, in the normal phase, all the conductivities
(i.e. all the entries of the conductivity matrix) can be expressed in terms
of a single, $\omega$-dependent function $f$. It is tempting then to interpret $f$
as a sort of mobility function for some would-be individual carriers. 
It should be noticed at once that the parametrization of the conductivity matrix
in terms of $f$ is made possible by the structure and symmetry of the equations of
motion for the fluctuations. The explicit form of the conductivity matrix in terms of
$f(\omega)$ is
\begin{eqnarray}\label{superrelazione}
\hat\sigma = \begin{pmatrix}  \sigma_A & \alpha T & \gamma \\ \alpha T &  \kappa T & \beta T \\ \gamma & \beta T & \sigma_B \end{pmatrix}=
\end{eqnarray}
\begin{eqnarray}
\begin{pmatrix} f \rho^2+1 & \frac{i \rho}{\omega}-\mu(f \rho^2+1)-\delta\mu f \rho\,\delta\rho & f \rho\ \delta\rho \\ \frac{i \rho}{\omega}-\mu(f \rho^2+1)-\delta\mu f \rho\,\delta\rho &  \kappa T &  \frac{i \delta \rho}{\omega}-\delta\mu(f \delta\rho^2+1)-\mu f \rho\,\delta\rho  \\ f \rho\ \delta \rho & \frac{i \delta \rho}{\omega}-\delta\mu(f \delta\rho^2+1)-\mu f \rho\,\delta\rho  & f \delta\rho^2+1 \end{pmatrix} \, .\nonumber
\end{eqnarray}

\begin{figure}[t]
\centering
\includegraphics[width=90mm]{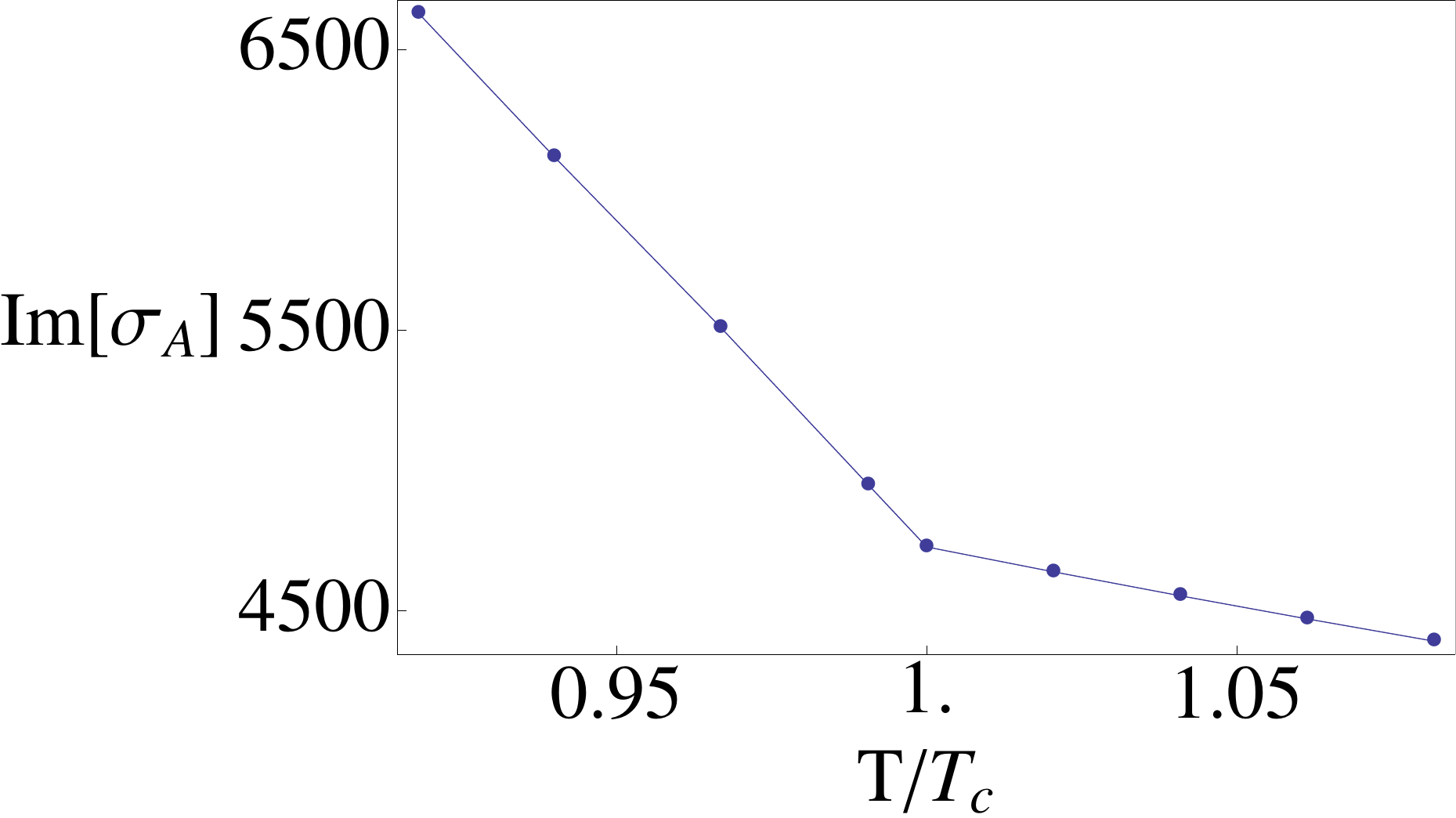} \vspace{1.5cm}
\\
\includegraphics[width=90mm]{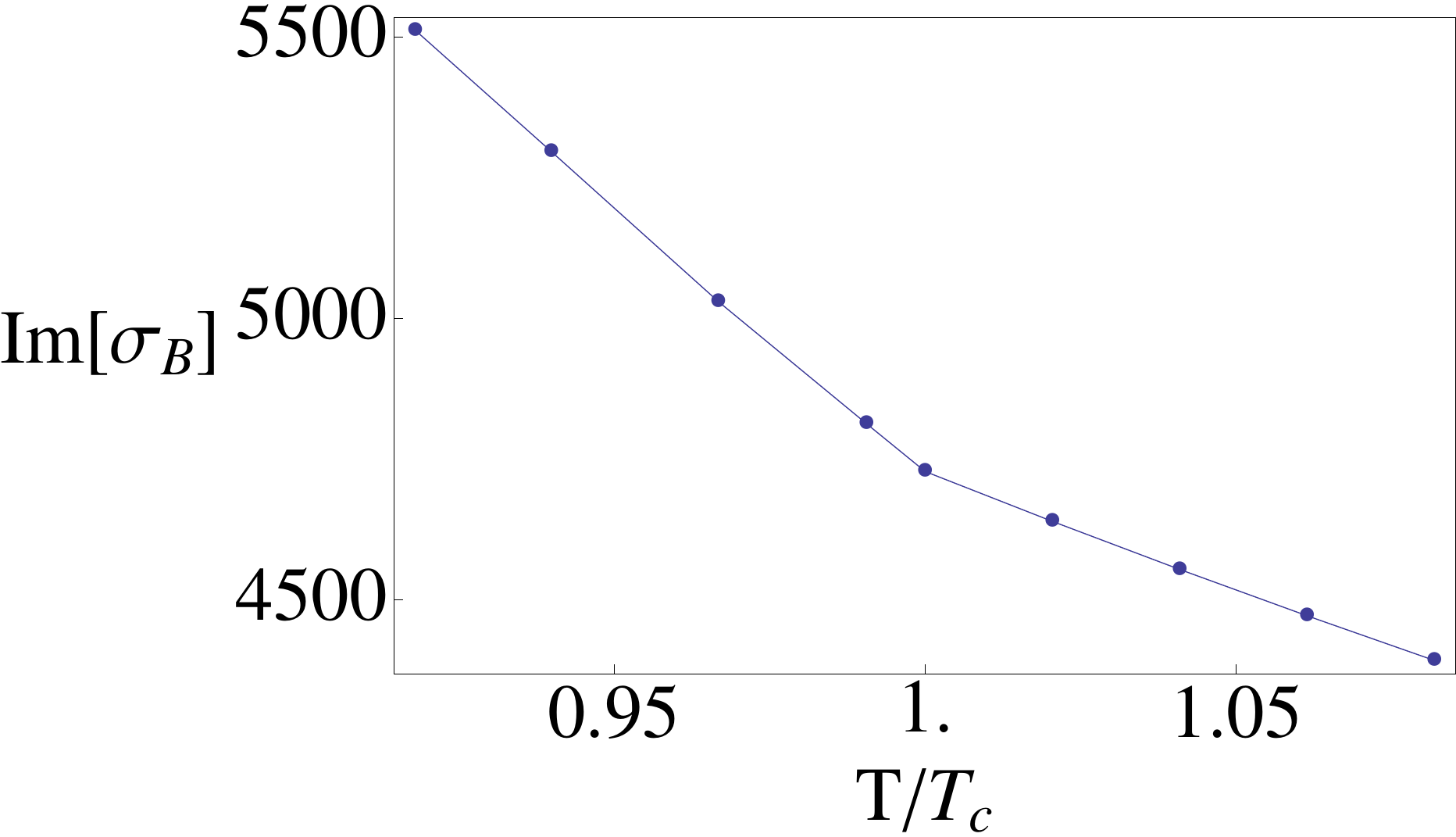}
\caption{Superconducting/normal state transition associated to
a discontinuous behavior of the derivative of the DC conductivity.
We have the ``electric'' and ``magnetic'' conductivities respectively in the 
upper and the lower plot. As described in the main text, notice that we have 
``spin superconductivity'' without spin symmetry breaking.}
\label{JUMPS}
\end{figure}

As our system is translationally invariant, both in the normal and in the superconducting phases,
there is no explicit source for momentum relaxation. We consequently have a diverging DC conductivity also in the normal phase;
this translates into the presence of a delta function in the real part of the conductivities at
$\omega=0$ also for $T>T_c$. Such phenomenon must not be confused with genuine superconductivity. In order to test the authentic
superconductivity of our holographic system for $T<T_c$, we have to be cautious and be able to separate the genuine superconductivity
from the simple diverging contribution due to translational invariance . In order to do so we have to study the amplitude 
of the DC delta function through the phase transition. Exploiting the Kramers-Kronig relations, the amplitude of
the DC delta function in the real part of a conductivity corresponds to the pole of the associated imaginary part
at $\omega=0$. We numerically study and plot this both for the electric and spin channels, see Figure \ref{JUMPS}.
We find then another very interesting result: the ``magnetic''
conductivity $\sigma_B$ shows a superconducting behavior below $T_c$.
This might sound surprising as the condensing field $\psi$
is charged only under the electric field $A$ and not $B$. 
Consequently, upon getting spontaneous and non-trivial 
VEV, $\psi$ breaks the electric symmetry and not the ``magnetic'' one. 
The ``spin-superconductivity'' manifested by our system is not 
directly due to the condensed degrees of freedom accounted for with $\psi$;
however the intertwined spin-electric properties described by the coupled
system of equations for the $A$ and $B$ gauge fluctuations
lead to a superconducting-like enhancement of DC spin transport below $T_c$.
Intuitively, an electric supercurrent flowing through our unbalanced
system affects the ``uncondensed'' spin degrees of freedom as well. 
This influence of the supercurrent flow on the spin degrees of freedom
is described holographically by the coupling of all the fields with the metric;
indeed the role of the backreaction is pivotal in determining the observed behavior.

\begin{figure}[t]
\centering
\includegraphics[width=70mm]{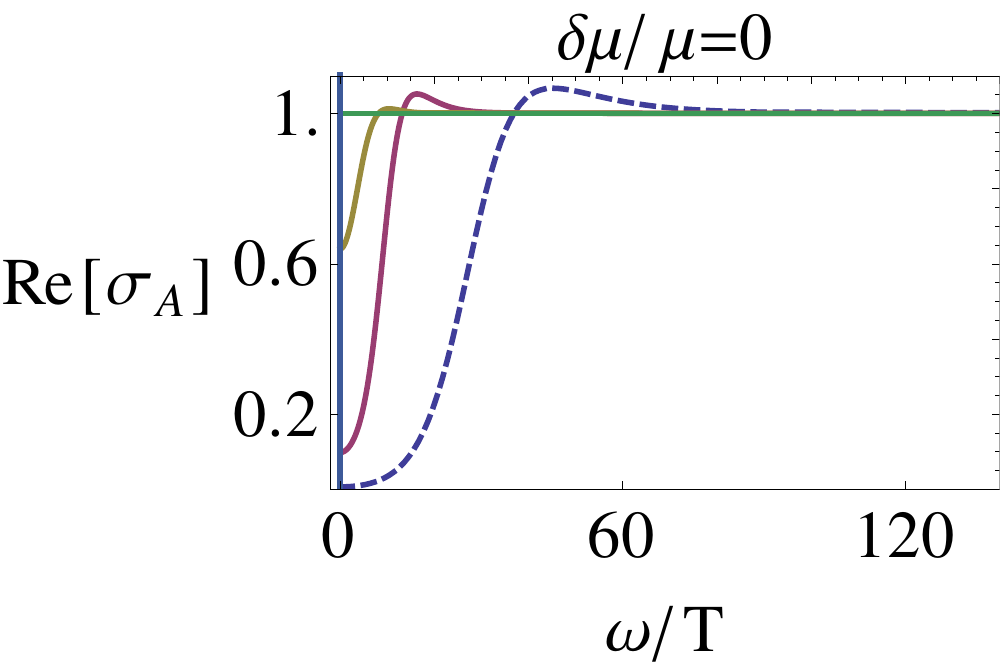} \hspace{0.5cm}
\includegraphics[width=70mm]{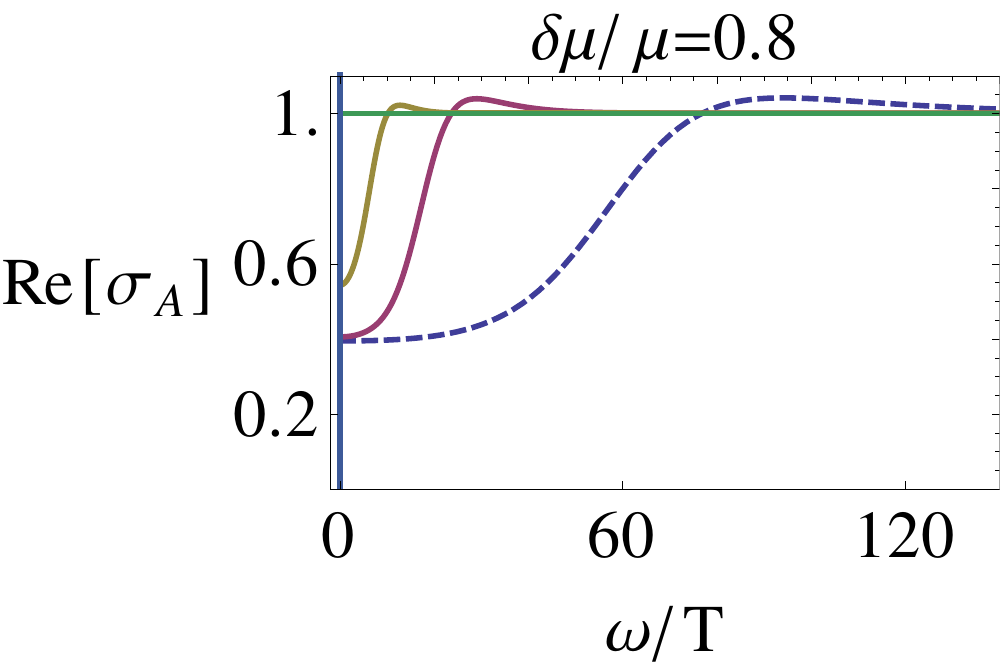}
\caption{Real part of the optical ``electric'' conductivity at various temperature $T>T_c$ 
in a balanced (left) and an unbalanced (right) situation. At high temperature
we obtain a featureless $\sigma_A$ (namely a horizontal line); for lower temperature, instead, a
depletion region at low $\omega$ becomes increasingly pronounced. The dashed lines 
correspond to $T=T_c$.}
\label{A}
\end{figure}

\begin{figure}[t]
\centering
\includegraphics[width=70mm]{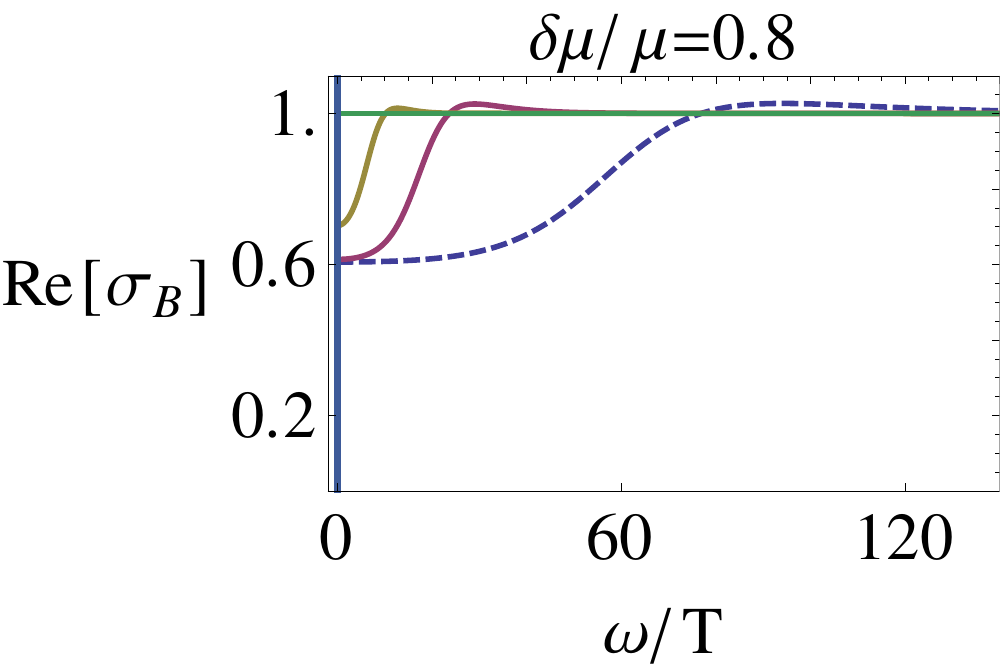} \hspace{0.5cm}
\includegraphics[width=70mm]{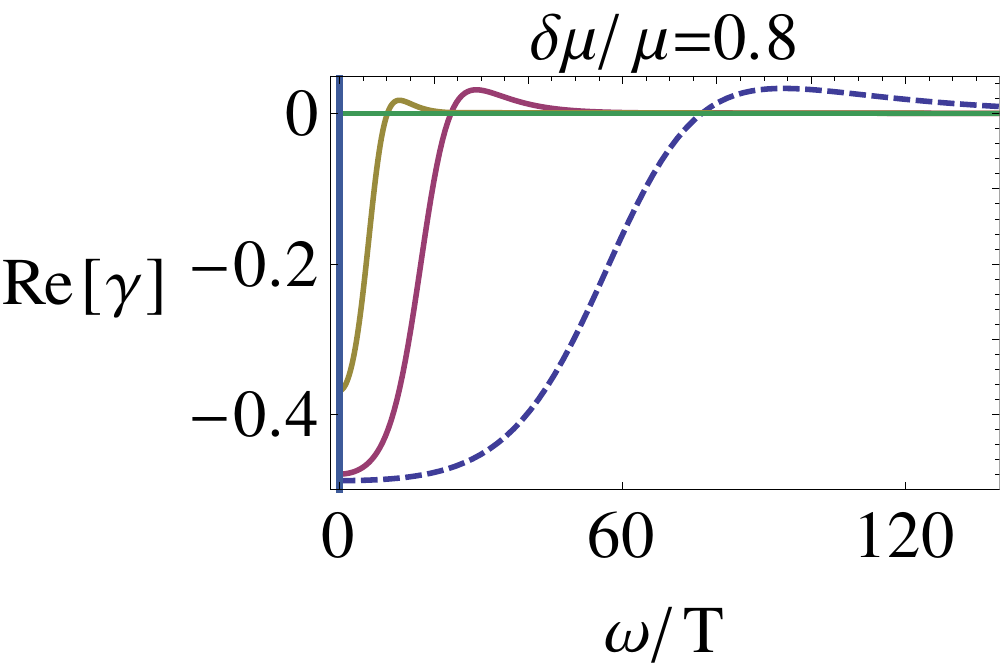}
\caption{Real part of the ``magnetic'' and mixed spin-electric conductivities in the
normal phase. The dashed line is associated to $T=T_c$.}
\label{BG}
\end{figure}

\begin{figure}[t]
\centering
\includegraphics[width=70mm]{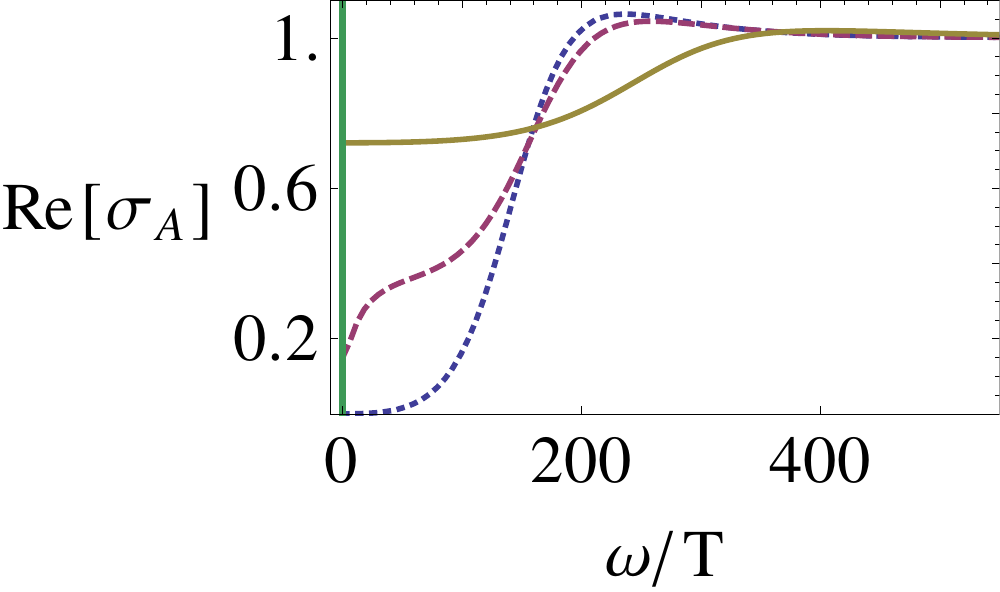} \hspace{0.5cm}
\includegraphics[width=70mm]{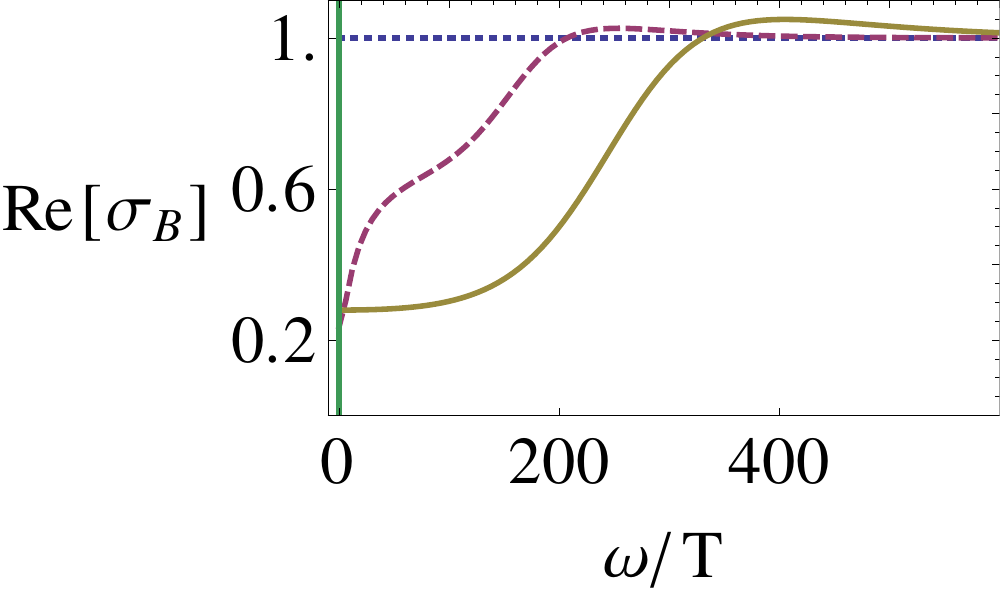}
\caption{The ``electric'' and ``magnetic'' conductivities in the 
superconducting phase. The features at small values of $\omega$ 
are due to the interplay of different scales; both $\mu$ and $\delta\mu$
are indeed different from zero.}
\label{SASB}
\end{figure}

\begin{figure}[t]
\centering
\includegraphics[width=70mm]{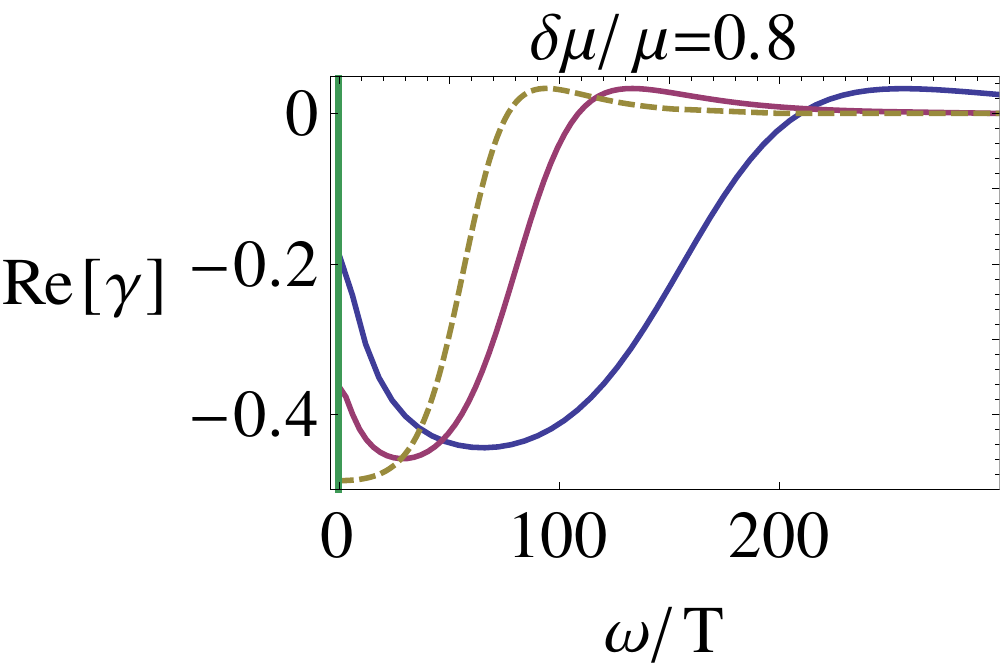} 
\caption{Mixed spin-electric conductivity in the superconducting phase for increasingly lower
temperature and fixed chemical imbalance. The dashed plot refers to the conductivity at $T_c$.}
\label{dude}
\end{figure}

\section{Conclusion}

We have here reported some results obtained in \cite{Bigazzi:2011ak} 
where a minimal model for an unbalanced holographic superconductor
was introduced and studied. In the present account we have added 
some new comments and observations especially about the interrelation between 
the chemical imbalance and the condensation, about the nature of the 
effective holographic approach to superconductivity and about
the time-reversal properties of the system.

The picture emerging at equilibrium shows that the chemical imbalance 
between the two species (interpreted here as ``spin-up'' and ``spin-down'' 
electrons) hinders the s-wave superconducting condensation;
this result matches the weak coupling expectation. Another result we obtained
is that our specific model does not manifest any Chandasekar-Clogston bound; 
this means that, at any value of the chemical imbalance, the system undergoes 
a superconducting phase transition at sufficiently low temperature. 
Such a feature emerging in our strongly coupled system is in contrast with the weak-coupling
expectation based on a BCS analysis. However, it has to be recognized that equilibrium properties
could be quite sensitive to the specific details of the model such as the values of
the mass and charge for the scalar field.  

The lack of a Chandrasekhar-Clogston bound suggests the absence of a LOFF phase
where an inhomogeneous superconducting condensate is thermodynamically favored
with respect to a homogeneous one. Such a conclusion is driven by the comparison of the
phase diagrams for the unbalanced superconductor at weak coupling, Figure \ref{weak_phase},
and the phase diagram of our strongly coupled model, Figure \ref{conde}.
In \cite{Bigazzi:2011ak} an analytical argument based on an analysis of the unbalanced 
superconductor in the so-called probe approximation was given
to exclude LOFF phases for our minimal model and parameter assignments.

The system at hand is very rich in relation to transport properties   and, in particular,
it shows interesting mixed features (i.e. effects related to the non-trivial off-diagonal entries in the
conductivity matrix). In its normal phase the systems can be regarded as a 
generalization to strong coupling of the simplest spintronic model (i.e. Mott model \cite{mott}).
A noteworthy feature is that here the optical conductivity matrix admits
a parametrization in terms of a single mobility function $f(\omega)$. This possibility
emerges directly from the structure of the equations of motion for the vector fluctuations
and its form is suggestively in line with a would-be quasi-particle-like interpretation.

We observed that the mixed spin-electric transport properties of our system are
crucially related to the coupling of the fields to the metric; in other words, to 
study mixed transport one needs to consider the backreaction of the fields on the geometry.
As a direct consequence we have that mixed transport phenomena are ubiquitous
in holographic systems whenever one consider the full backreacted gravitational system.
More generally, let us mention that, as opposed to the equilibrium 
properties of the system, its transport features are universal and insensitive to 
the details of the model such as the values of the scalar field parameters.

\section{Future perspectives}

The present system allows for many generalizations; one among the simplest
extensions consists in considering a second scalar field which could serve
as an order parameter for the ``magnetic'' $U(1)$. Such possibility has been 
already investigated in \cite{Musso:2013ija} relying on a probe-approximation
analysis. Such extended system is interesting in view of a holographic exploration
of the coexistence of different order parameters and their mutual interaction
(e.g. competition/enhancement).

The mixed spin-electric conductivity in the superconducting phase shows an interesting
enhancement in the low $\omega$ region. In a recent paper, \cite{Amado:2013xya}, a similar
feature has been argued to be possibly related to momentum relaxation and Drude-like behavior.
This interesting possibility has to be further investigated and the present model
could offer a simple playground on which similar ideas could be tested.

It recently appeared a paper studying angular momentum and spin transport
relying on the analysis of the dual bulk spin connection \cite{Hashimoto:2013bna}. It would be interesting
to understand the relations between their and our approach to address spin transport in 
a holographic context.

\section{Acknowledgements}

I am extremely grateful to Francesco Bigazzi, Aldo Cotrone, Alberto Lerda, Davide Forcella, Riccardo Argurio, Johanna Erdmenger,
Lorenzo Calibbi, Daniel Arean, Ignacio Salazar Landea, Massimo D'Elia, Carlo Maria Becchi, Alessandro Braggio, 
Andrea Amoretti, Diego Redigolo, Nicola Maggiore, Nicodemo Magnoli, Katherine Michele Deck and Hongbao Zhang for their useful advices and 
very interesting conversations. My work was partially supported by the ERC Advanced Grant
"SyDuGraM", by IISN-Belgium (convention 4.4514.08) and by the
``Communaut\'e Fran\c{c}aise de Belgique" through the ARC program.

\end{document}